# Genuine and Matter-induced Components of the CPV Asymmetry for Neutrino Oscillations

**J. Bernabéu[1]**

*Department of Theoretical Physics, University of Valencia, and IFIC, U.Valencia-CSIC*
*E-46100 Burjassot, Valencia*
*E-mail:* `jose.bernabeu@uv.es`

**A. Segarra**

*Department of Theoretical Physics, University of Valencia, and IFIC, U.Valencia-CSIC*
*E-46100 Burjassot, Valencia*
*E-mail:* `alejandro.segarra@uv.es`

These results represent the solution for the historical problem of the contamination by matter effects on the CPV Asymmetry for neutrino oscillations. Vacuum is CPT-symmetric and matter is T-symmetric, the goal is accomplished by using this guiding principle. Independent of the theoretical framework for the dynamics of the active neutrino flavors, we prove the Disentanglement Theorem A(CP)=A(CP, T)+A(CP, CPT) for the experimental CPV Asymmetry, with A(CP, T) genuine T-odd and A(CP, CPT) fake CPT-odd.

For the effective Hamiltonian written as the sum of free mass propagation plus the matter potential for electron-neutrinos, the two components have definite parities under the baseline L, the matter potential "$a$", the imaginary part sin$\delta$ of the PMNS mixing matrix and the hierarchy "h"=±1 in the neutrino mass ordering: A(CP, T) is odd in L and sin$\delta$ plus even in $a$ and h; A(CP, CPT) is even in L and sin$\delta$ plus odd in $a$ and almost odd in h.

For present terrestrial accelerator sources of muon-neutrinos and antineutrinos, the two components of the appearance CPV asymmetry A(CP) can be disentangled by either baseline dependence (HKK) or energy dependence (DUNE). At the DUNE baseline, the higher energy region above the first oscillation node provides a dominant matter-induced A(CP, CPT) component and the sign of the experimental asymmetry A(CP) gives the hierarchy in the neutrino mass ordering. On the contrary, there is a "magic energy" E around the second oscillation maximum in which the fake A(CP, CPT) component has a first-rank zero whereas the genuine A(CP, T) component has a maximum proportional to sin$\delta$. With a modest energy resolution $\Delta E \sim 200$ MeV an effective zero remains in the matter-induced A(CP, CPT).

---

[1]Speaker

1.   Introduction

The prehistory of neutrino physics ended in 1998 with the discovery of neutrino oscillations in atmospheric neutrinos [1], also observed later in solar neutrinos [2] and in reactor and accelerator experiments. Since then, a consistent dynamical framework of neutrino masses and flavor mixing has been established. Values of the three mixing angles for the three active neutrino species and the two mass differences $\Delta m_{21}^2$ and $|\Delta m_{31}^2|$ have been fixed. Open questions are the neutrino Dirac/Majorana nature, the absolute mass scale, the hierarchy in the neutrino mass ordering and CP-Violation. The first two of these properties are not accessible by neutrino flavor oscillations. The study presented in this paper affects the last two properties.

In underground neutrino propagation from the source to the detector, the dynamics is affected by the interaction with matter. The MSW [3] effect of charged-current weak interactions (V) of electron-neutrinos with matter electrons adds a new term in the effective Hamiltonian for flavor oscillations to the vacuum term induced by the PMNS [4] mixing matrix (*U*) plus neutrino mass differences. In the flavor basis, with $a = 2E.V$,

$$H = \frac{1}{2E}\left\{U\begin{bmatrix} m_1^2 & 0 & 0 \\ 0 & m_2^2 & 0 \\ 0 & 0 & m_3^2 \end{bmatrix}U^\dagger + \begin{bmatrix} a & 0 & 0 \\ 0 & 0 & 0 \\ 0 & 0 & 0 \end{bmatrix}\right\} \qquad (1)$$

for neutrinos. For antineutrinos, U → U*, $a$ → -$a$. As seen in Eq. (1), CP-Violation (CPV) has now two different origins: (i) genuine in the complexity of U(PMNS); (ii) the CP-asymmetry of the Earth, "$a$" changing sign from left-handed neutrinos to right-handed antineutrinos. This second fake component has been a historical problem which was not disentangled from the first genuine component. A direct evidence of Symmetry Violation, which means the measurement in a single experiment of a genuine Observable Odd under the Symmetry, should be however a fundamental objective.

How to search for genuine CPV? Its interest is evident in itself, as a symmetry in the laws of physics, and also for giving an opportunity to Leptogenesis [5] for the Matter-Antimatter Asymmetry in the Universe. The community seems to have accepted the historical problem as irremediable and global fits (or single experiment fit in the future) for observables calculated with the Hamiltonian H are made in order to extract the δ phase of U(PMNS) in the vacuum term. This methodology is potentially dangerous, not only because it is theory-dependent, because you may obtain δ ≠ 0, π without any contribution of genuine CPV terms in your observables (!). In the transition probabilities, the δ-dependent terms appear with their behavior under the three CP, T, CPT symmetries as shown in Table 1

|  | CPTI | CPTV |
|---|---|---|
| CPV | $\sin\delta$ | $a$,   $a\cos\delta$ |
| CPC | $\cos\delta$ | $a\sin\delta$ |

Arrows: upper-left and lower-left arrows point outward; upper-right → TRI; lower-right → TRV.

Table 1

For T-symmetric matter, $a$ is odd under CP and CPT, hence its influence on the fake CPTV second column of Table 1. Only the term $\sin\delta$ corresponds to genuine CPV, the first column is CPT-invariant (CPTI) as in vacuum, the first row is genuine and fake CPV whereas the entire second row is CP-conserving (CPC). The diagonal terms are TRV, whereas the antidiagonal terms are T-invariant(TRI). As you will recognize later, present fits (hints) for δ-dependence demonstrate that the genuine CPV term in the intervening observables plays, if any, a minor role.

In Section 2 we discuss the theoretical solution of the historical problem by means of an **Asymmetry Disentanglement Theorem** [6] which gives the asymmetry in terms of two components with opposite behavior under CPT. For the transition channel $\nu_\mu \rightarrow \nu_e$ the two components of the CPV-asymmetry are calculated in terms of the parameters of the Hamiltonian. In section 3 the experimental signatures [7] for the separation of the two components are studied, with complementary strategies for the two planned experiments HK [8] and DUNE [9]. Section 4 summarizes our conclusions.

## 2. Disentangling Genuine and Fake CPV Components

The guiding principle for the disentanglement of genuine and fake components of the CPV asymmetry is their behavior under the three CP, T, CPT symmetries. Whereas both vacuum and matter-induced terms are CPV, the first is CPTI, the second is TRI. The conceptual basis, independent of the theoretical framework for describing the dynamics of neutrino propagation in matter, is to write the effective Hamiltonian as

$$H = \frac{1}{2E}\widetilde{U}\widetilde{M}^2\widetilde{U}^\dagger \qquad (2)$$

connecting flavor with definite mass in matter $\widetilde{m}$ by means of $\widetilde{U}$. The Observables in matter depend on the rephasing-invariant mixings $\widetilde{J}^{ij}_{\alpha\beta} \equiv \widetilde{U}_{\alpha i}\widetilde{U}^*_{\alpha j}\widetilde{U}^*_{\beta i}\widetilde{U}_{\beta j}$ and the oscillation phases $\widetilde{\Delta}_{ij} \equiv \frac{\Delta\widetilde{m}^2_{ij}L}{4E}$ where $\widetilde{J}^{ij}_{\alpha\beta}$ and $\widetilde{\Delta}_{ij}$ are both energy-dependent.

From the different behavior of these last ingredients under the discrete T and CPT symmetry transformations one derives [6] the **Asymmetry Disentanglement Theorem** by separating the observable CPV-asymmetry in any flavor transition α → β into L-even (TRI, CPTV) and L-odd (TRV, CPTI) components, where L is the Baseline,

$$A_{\alpha\beta}^{CP} = A_{\alpha\beta}^{CPT} + A_{\alpha\beta}^{T},$$

T-invariant ⤶          ⤷ CPT-invariant

$$A_{\alpha\beta}^{CPT} = -4 \sum_{j<i} \left[ Re\tilde{J}_{\alpha\beta}^{ij} \sin^2\tilde{\Delta}_{ij} - Re\tilde{\bar{J}}_{\alpha\beta}^{ij} \sin^2\tilde{\bar{\Delta}}_{ij} \right]$$

$$A_{\alpha\beta}^{T} = -2 \sum_{j<i} \left[ Im\tilde{J}_{\alpha\beta}^{ij} \sin 2\tilde{\Delta}_{ij} - Im\tilde{\bar{J}}_{\alpha\beta}^{ij} \sin 2\tilde{\bar{\Delta}}_{ij} \right]$$

(3)

where $\bar{J}$ and $\bar{\Delta}$ refer to antineutrinos. These Eqs. (3) remain valid for extended neutrino models with more mass eigenstates and a rectangular mixing matrix, not only for the three neutrino model.

The two components have definite parities under L -model independent-, $a$, sin δ and the hierarchy h= ±, in the neutrino mass ordering: even h means blind to the hierarchy, odd h means changing sign with the change of hierarchy. A(CP, T) is CPTI, odd in L and sinδ plus even in $a$ and h(!!!). The fact that the genuine CPV component is blind to h is in contrast with the result for δ obtained in global fits. A(CP, CPT) is TRI, even in L and sinδ plus odd in $a$ and almost odd in h.

This behavior is explicitly seen in analytic approximations for the two components which are valid for actual HK and DUNE experiments. In the energy region between the two MSW resonances $\Delta m_{21}^2 \ll a \ll |\Delta m_{31}^2|$ one obtains [7]

$$A_{\mu e}^{CPT} = 16A\left(\frac{\sin\Delta_{31}}{\Delta_{31}} - \cos\Delta_{31}\right)(S \sin\Delta_{31} + J_r \cos\delta \Delta_{21} \cos\Delta_{31}) + O(A^3),$$

$$A_{\mu e}^{T} = 16 J_r \sin\delta\, \Delta_{21} \sin^2\Delta_{31} + O(A^2)$$

(4)

where $S \equiv c_{13}^2 s_{13}^2 s_{23}^2$, $J_r \equiv c_{12} c_{13}^2 c_{23} s_{12} s_{13} s_{23}$, $A \equiv \frac{aL}{4E} \propto L$ and $\Delta_{ij} \equiv \frac{\Delta m_{ij}^2 L}{4E} \propto L/E$

As indicated in Eqs. (4), the corrections in "A" are quadratic.

3.     Experimentals Signatures: Magic Energy

The two planned detectors HK and DUNE offer different characteristics in baseline, type of neutrino beam and neutrino energy identification, so the strategies to be followed for an experimental separation of the genuine and fake components of the CPV asymmetry appear distinct.

In the case of HK we have an Off-Axis beam of $v_\mu$ neutrinos adressed to Kanioka with L = 295 Km. The corresponding 1st. oscillation maximum is at 0,6 GeV. It can be selected with an Off-Axis angle $\theta_{oa} = 2,7°$. The complete neutrino spectrum is above the first oscillation node and we give in Fig. 1 the two components as function of neutrino energy for both Normal Hierarchy(NH) and Inverted Hierarchy(IH) for the νμ $v_\mu \rightarrow v_e$ transition

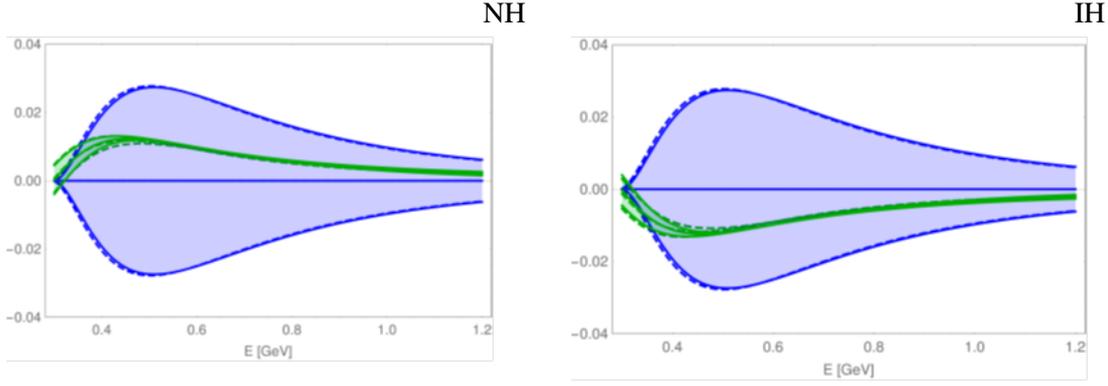

NH    IH

Fig. 1

The blue band is for the genuine component and the green band is for the fake component. They cover all values of the CPV phase δ = 0 to 2π: whereas A(CP,T) is proportional to sin δ, A(CP, CPT) has a minor dependence on cos δ. As anticipated, the genuine component is blind to the hierarchy, whereas the sign of the fake component identifies the hierarchy. At this baseline, matter-effects are small and they can be subtracted in order to reconstruct the genuine component of the asymmetry from the experiment. A separation strategy could be envisaged with two detectors HKK, the second detector in Korea at L = 1000-1300 Km with the central spectrum at 0° peaking in the first oscillation maximum at 2.1-2.7 GeV. The relative contribution of A(CP, CPT) increases with L, then dominates at these large energies and its sign determines the hierarchy. The two components would thus be separated by baseline-dependence. If these high energies were not accessible, there is a second alternative: the second oscillation maximum at L = 1000-1300 Km. peaks at 0.70-0.92 GeV, selected with an Off-Axis beam at $\theta_{0a}$= 1.9-2.4$^0$. The results for the genuine and fake components are given for L = 1300 Km in Fig. 2 as function of neutrino energy

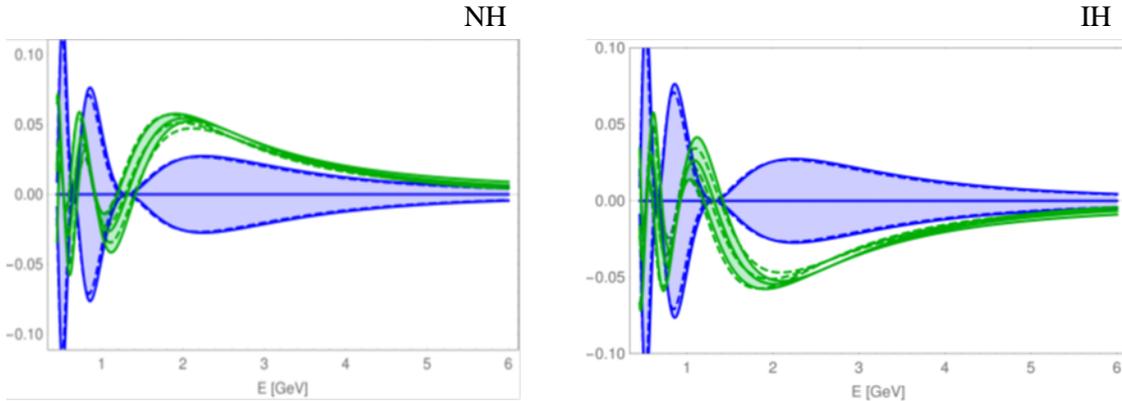

NH    IH

Fig. 2

We discover the existence of a **Magic Energy** providing a direct measurement of the genuine A(CP, T) component near the second oscillation maximum at 0.92 GeV. There the fake A(CP, CPT) component presents a first-rank zero, independent of δ and h.

In the case of DUNE at L = 1300 Km. we have a wide-band beam and the energy is reconstructed from the detector. The strategy for separation is feasible by measuring the CPV asymmetry as function of E-bins. The zero of A(CP, CPT) at the **Magic Energy** being of first-rank helps in needing a modest energy-resolution only. For E = 0.92 ± 0.15 GeV, the experimental asymmetry measures the genuine component and it is proportional to sin δ. At large energies, the sign of the experimental asymmetry measures the hierachy independent of δ.

We may understand the simultaneous existence of the first-rank zero of the fake component and the maximum of the genuine component around the second oscillation maximum from the analytic expresions of Eqs. (4). One obtains

$$\text{Maxima of A(CP, T)} \rightarrow \tan \Delta_{31} = -2 \Delta_{31} \quad (5.a)$$

$$\delta\text{-independent zeros of A(CP, CPT)} \rightarrow \tan \Delta_{31} = \Delta_{31} \quad (5.b)$$

The solutions of (5.a) and (5.b) include, respectively, values nearly above and below the second oscillation maximum at $\Delta_{31} = 3\pi/2$. Explicitly, the first-rank zero of A(CP, CPT) in (5.b) determines the Magic Energy as

$$E = 0.92 \; GeV \; \frac{L}{1300 \; km} \frac{|\Delta m^2_{31}|}{2.5*10^{-3} \; eV^2} \quad (6)$$

only dependent on $|\Delta m^2_{31}|$. Using its present fitted value [10] we obtain L/E = 1420 km/GeV for the correlation between the baseline and the magic energy.

4.      Conclusion

The Physics of the CPV Asymmetry for Neutrino Oscillations in Matter is now understood in terms of two components: a genuine CP and T-odd term and a matter-induced CP and CPT-odd term. They have definite and opposite parities under L -theory independent-, CPV sin δ, matter-potential $a$ and hierarchy h.

With muon neutrino and antineutrino sources only, the experimental CPV-Asymmetry for the appearance transition $\nu_\mu \rightarrow \nu_e$ can separate the genuine and fake components by either Baseline-dependence or Energy-dependence. The genuine A(CP, T) component is blind to the hierarchy h and it is proportional to sinδ. The fake A(CP, CPT) component is almost odd in h and almost independent of δ.

The fake component A(CP, CPT) vanishes at the Magic Energy, independent of h, δ and $a$; only $|\Delta m^2_{31}|$ matters. It appears near the second oscillation maximum where the genuine A(CP, T) component is also maximal. The conditions for a direct evidence of genuine CP Violation in neutrino oscillations in matter are thus met.

ACKNOWLEDGEMENTS
This research has been supported by MINECO Project FPA 2017-84543-P, Generalitat Valenciana Project GV PROMETEO 2017-033 and Severo Ochoa Excellence Centre Project SEV 2014-0398. A.S. acknowledges the MECD support through the FPU14/04678 grant.

# References

[1] Y. Fukuda et al. [Super-Kamiokande Collaboration], *Evidence for oscillation of atmospheric neutrinos*, Phys. Rev. Lett. 81 (1998) 1562–1567

[2] Q. R. Ahmad et al. [SNO Collaboration], *Direct evidence for neutrino flavor transformation from neutral current interactions in the Sudbury Neutrino Observatory*, Phys. Rev. Lett. **89** (2002) 011301.

[3] L. Wolfenstein, *Neutrino Oscillations in Matter, Phys*. Rev. **D17** (1978) 2369–2374

S. P. Mikheyev and A. Yu. Smirnov, *Resonance Amplification of Oscillations in Matter and Spectroscopy of Solar Neutrinos*, Sov. J. Nucl. Phys. **42** (1985) 913–917

[4] B. Pontecorvo, *Mesonium and anti-mesonium*, Sov. Phys. JETP **6** (1957) 429, [Zh. Eksp. Teor. Fiz. **33** (1957) 549].

Z. Maki, M. Nakagawa, and S. Sakata, *Remarks on the unified model of elementary particles*, Prog. Theor. Phys. **28** (1962) 870–880.

[5] M. Fukugita and T. Yanagida, *Baryogenesis Without Grand Unification*, Phys. Lett. **B174** (1986) 45–47.

[6] J. Bernabéu and A. Segarra, *Disentangling genuine from matter-induced CP violation in neutrino oscillations*, Phys. Rev. Lett. **121**, no. 21 (2018) 211802.

[7] J. Bernabéu and A. Segarra, *Signatures of the genuine and matter-induced components of the CP violation asymmetry in neutrino oscillations*, JHEP **11** (2018) 063.

[8] A K. Abe et al. [Hyper-Kamiokande Collaboration], *Hyper-Kamiokande Design Report* (2018). arXiv: 1805.04163 [physics.ins-det].

[9] A R. Acciarri et al. [DUNE Collaboration], Long-Baseline Neutrino Facility (LBNF) and Deep Underground Neutrino Experiment (DUNE) (2015). arXiv: 1512.06148 [physics.ins-det].

[10] A P. F. de Salas, D. V. Forero, C. A. Ternes, M. Tortola, and J. W. F. Valle, *Status of neutrino oscillations 2018: 3s hint for normal mass ordering and improved CP sensitivity*, Phys. Lett. **B782** (2018) 633–640.